\documentclass[aps,nofootinbib,twocolumn,groupbyaffilation,showpacs,floatfix,longbibliography]{revtex4-1}

\usepackage{graphicx}
\usepackage{dcolumn}
\usepackage{bm}
\usepackage{amssymb}
\usepackage{microtype}
\usepackage{xfrac}
\usepackage{gensymb}
\usepackage{xcolor}
\usepackage[percent]{overpic} 
 
\usepackage{natbib,hyperref}
\hypersetup{
	colorlinks=true, 
	linktoc=section,     
	linkcolor=blue,  
}

\begin{document}

\title{Frustrated Magnetism in Fluoride and Chalcogenide Pyrochlore Lattice Materials}

\author{Dalmau Reig-i-Plessis}
\affiliation{Department of Physics and Astronomy and Quantum Matter Institute, University of British Columbia, Vancouver, British Columbia V6T 1Z1, Canada}

\author{Alannah~M.~Hallas}
\affiliation{Department of Physics and Astronomy and Quantum Matter Institute, University of British Columbia, Vancouver, British Columbia V6T 1Z1, Canada}
\email[Email: ]{alannah.hallas@ubc.ca}

\date{\today}

\begin{abstract}
Pyrochlore lattices, which are found in two important classes of materials -- the $A_2B_2X_7$ pyrochlore family and the $AB_2X_4$ spinel family -- are the quintessential 3-dimensional frustrated lattice architecture.
While historically oxides ($X =$~O) have played the starring role in this field, the past decade has seen materials synthesis breakthroughs that have lead to the emergence of fluoride ($X =$~F) and chalcogenide ($X =$~S, Se) pyrochlore lattice materials.
In this Research Update, we summarize recent progress in understanding the magnetically frustrated ground states in three families of non-oxide pyrochlore lattice materials: (i) $3d$-transition metal fluoride pyrochlores, (ii) rare earth chalcogenide spinels, and (iii) chromium chalcogenide spinels with a breathing pyrochlore lattice. 
We highlight how the change of anion can modify the single ion spin anisotropy due to crystal electric field effects, stabilize the incorporation of new magnetic elements, and dramatically alter the exchange pathways and thereby lead to new magnetic ground states.
We also consider a range of future directions -- materials with the potential to define the next decade of research in frustrated magnetism. 
\end{abstract}

\maketitle

\section{Introduction} \label{intro}

In the frustrated magnetism community, interest in pyrochlore oxides flourished following the 1997 discovery of the spin ice state in Ho$_2$Ti$_2$O$_7$~\cite{harris1997geometrical,ramirez1999zero}, and shortly thereafter in 1999, the discovery of a putative spin liquid state in Tb$_2$Ti$_2$O$_7$~\cite{gardner1999cooperative}. 
Since that time, a detailed characterization of the magnetic ground state of almost every known oxide pyrochlore has been undertaken~\cite{10_Gardner_Magnetic_pyrochlore_oxides}, including in recent years, pyrochlores that can only be formed under high pressure conditions~\cite{wiebe2015frustration}.
Furthermore, the magnetic ground states of these pyrochlores have been poked and prodded in every imaginable way, from extreme applied pressure~\cite{mirebeau2002pressure} to high magnetic fields~\cite{ross2011quantum} to exacting studies of the effect of off-stoichiometry~\cite{arpino2017impact} and most recently their growth as epitaxial thin films~\cite{bovo2014restoration}.
This enormous body of research has revealed that the magnetic ground states of many oxide pyrochlores are exceedingly fragile, with very few materials sharing the same magnetic phenomenology.
The origin of these diverse magnetic states is, first and foremost, the frustrated lattice architecture of the corner-sharing tetrahedral pyrochlore lattice but the second most important factor is the nature of the magnetic ion itself, both its single ion properties and its interactions with neighbors. 

Pyrochlore oxides, with the general formula $A_2B_2$O$_7$, take their name from a naturally occurring mineral.
This name is also bestowed upon the lattice of corner-sharing tetrahedra, known as the pyrochlore lattice, which both the $A$ and $B$ atoms form in the eponymous family of materials.
Pyrochlore oxides have extensive chemical versatility~\cite{subramanian1983oxide}, which is another factor that contributes to the diversity in their magnetic states. 
The spinel oxides, $AB_2$O$_4$, which also take their name from a naturally occurring mineral, provide a second structural platform to study frustrated magnetism on the pyrochlore lattice. 
However, in the spinel structure, only  $B$ forms a pyrochlore lattice while $A$ makes up a diamond lattice. 
Like the pyrochlore oxides, spinel oxides have a significant chemical versatility, yielding a large number of compounds~\cite{hill1979systematics,sickafus1999structure}. 
However, many of these materials have magnetic $A$ sites, which spoils the simple frustration of the pyrochlore $B$ site.
Nonetheless, there are many compounds in the spinel family with geometric-frustration driven physics due to both orbital and magnetic degrees of freedom~\cite{10_SH_Lee_spinels,03_Tsunetsugu_vanadium_spinels}.

While oxide pyrochlore lattice materials have dominated the field for multiple decades, recent advances in materials synthesis have brought several non-oxide families to the forefront, including fluoride pyrochlores and chalcogenide spinels.
The chemical and structural modifications that accompany this replacement of the anion in turn leads to different local environments and hence different single ion anisotropies, different relative energy scales, and new magnetic ions on the pyrochlore lattice, all of which have lead to significant new physics.

In this Research Update we outline what's been accomplished with nonoxide pyrochlores, primarily over the past five years, and discuss challenges and opportunities for the future. 
In Section~\ref{fluoride}, we introduce the fluoride pyrochlores which have the unique characteristic of hosting late $3d$ magnetic transition metals, which cannot be achieved in oxide pyrochlores.
In Section~\ref{RE-spinels}, we discuss the rare earth chalcogenide spinels, which have distinct properties from their oxide pyrochlore counterparts due to the different local environment and the reduced crystal field energy scale. 
In Section~\ref{Cr-spinel}, we turn to the chromium chalcogenide spinels with an emphasis on the breathing pyrochlore spinels that have alternating small and large tetrahedra, breaking the degeneracy of the pristine pyrochlore lattice but unlocking new, exotic magnetic states.
The magnetic properties of the materials discussed in Sec.~\ref{fluoride} - \ref{Cr-spinel} are summarized in Table~\ref{table:compound_summary}.
Finally, in Section~\ref{future}, we turn to future directions:  known materials that as of now have been the subjects of relatively few investigations or places where altogether unknown, new materials might be lurking.

\begin{figure*}
    \centering
    \includegraphics[width=7in]{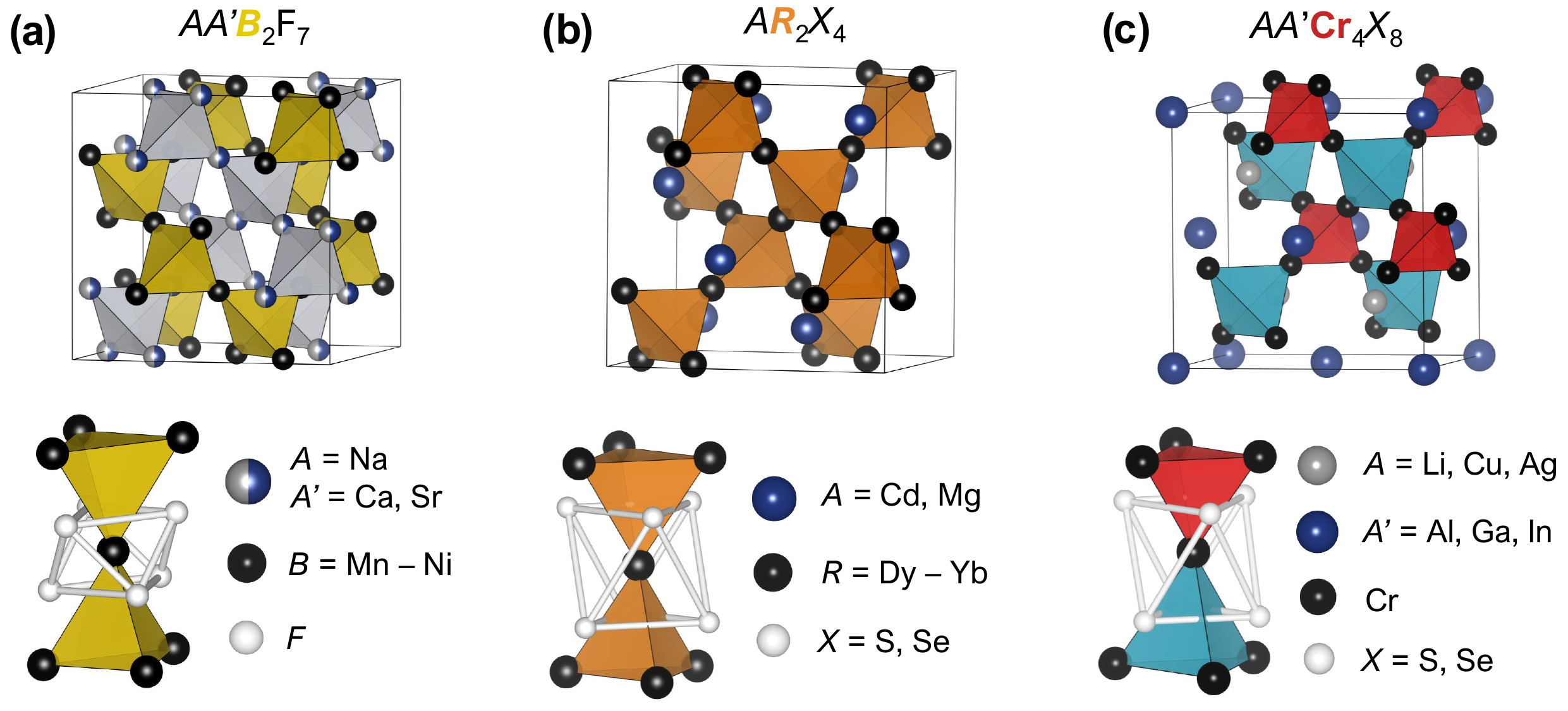}
    \caption{The crystal structures of three families of non-oxide materials with pyrochlore (corner sharing tetrahedral) sublattices. (a) The fluoride pyrochlores, $AA'B_2$F$_7$, where the $B$-site pyrochlore sublattice, highlighted in yellow, is occupied by a $3d$ transition metal. (b) The rare earth chalcogenide spinels, $AR_2X_4$, where the $R$-site pyrochlore sublattice, highlighted in orange, is occupied by a rare earth. (c) The breathing pyrochlore chromium spinels, $AA'$Cr$_4X_8$, where the Cr pyrochlore sublattice has alternating small (red) and large (teal) tetrahedra. The top set of panels show the unit cell for each structure and the lower set of panels show the local anion environment at the magnetic site.}
    \label{fig:structures}
\end{figure*}

\begin{table}[htbp]
\caption{Crystal structure details, including the Wyckoff site (Wyck.), point group symmetry (P.G.), atomic coordinates ($x$,~$y$,~$z$), and oxidation state (Ox.), for the fluoride pyrochlores, $AA'B_2$F$_7$, the chalcogenide spinels, $AR_2X_4$, and the breathing pyrochlore chalcogenides, $AA'$Cr$_4X_8$.}
\label{table:crystal_structures}
\begin{ruledtabular}
\begin{tabular}{l c c c c c c}
\multicolumn{7}{c}{\textbf{Fluoride Pyrochlores} $AA'B_2$F$_7$} \\ 
\multicolumn{7}{c}{Space Group Fd$\bar{3}$m}                \\
\hline
       & Wyck.  & P.G.      & $x$   & $y$   & $z$   & Ox.   \\ 
\hline
$A/A'$ & $16d$  & $D_{3d}$  & 0.5   & 0.5   & 0.5   & +1/+2 \\
$B$    & $16c$  & $D_{3d}$  & 0     & 0     & 0     & +2    \\
F1     & $8b$   & $T_d$     & 0.375 & 0.375 & 0.375 & $-1$  \\
F2     & $48f$  & $C_{2v}$  & $x$   & 0.125 & 0.125 & $-1$  \\ 
\hline
       &        &           &       &       &       &       \\ 
\multicolumn{7}{c}{\textbf{Chalcogenide Spinels} $AR_2X_4$} \\
\multicolumn{7}{c}{Space Group Fd$\bar{3}$m}                \\ 
\hline
        & Wyck. & P.G.      & $x$   & $y$   & $z$   & Ox.   \\ 
\hline
$A$     & $8a$  & $T_d$     & 0.125 & 0.125 & 0.125 & +2    \\
$R $    & $16d$ & $D_{3d}$  & 0.5   & 0.5   & 0.5   & +3    \\
$X $    & $32e$ & $C_{3v}$  & $x$   & $x$   & $x$   & $-2$  \\ 
\hline
       &        &           &       &       &       &       \\ 
\multicolumn{7}{c}{\textbf{Breathing Pyrochlores} $AA'$Cr$_4X_8$} \\ 
\multicolumn{7}{c}{Space Group F$\bar{4}$3m}                \\ 
\hline
        & Wyck. & P.G.      & $x$   & $y$   & $z$   & Ox.   \\ 
\hline
$A $    & $4c$  & $T_d$     & 0.25  & 0.25  & 0.25  & +1    \\
$A'$    & $4a$  & $T_d$     & 0     & 0     & 0     & +3    \\
Cr      & $16e$ & $C_{3v}$  & $x_1$ & $x_1$ & $x_1$ & +3    \\
$X1 $   & $16e$ & $C_{3v}$  & $x_2$ & $x_2$ & $x_2$ & $-2$  \\
$X2 $   & $16e$ & $C_{3v}$  & $x_3$ & $x_3$ & $x_3$ & $-2$  \\ 
\end{tabular}
\end{ruledtabular}
\end{table}

\section{Fluoride Pyrochlores}\label{fluoride}

While the late $3d$ transition metals, from Fe to Cu, are among the most important elements in studies of quantum magnetism, one may lament that they are entirely unrepresented among the pyrochlore oxides, $A_2B_2$O$_7$. 
The reason for this is that charge neutrality and structural tolerance factors cannot be simultaneously satisfied for any combination of $A$ and $B$ cations that includes a late $3d$ transition metal~\cite{subramanian1983oxide}. 
In recent years, this obstacle has been sidestepped by replacing divalent oxygen, O$^{2-}$ with monovalent fluorine, F$^-$, allowing pyrochlores with $B=$~Mn$^{2+}$, Fe$^{2+}$, Co$^{2+}$ and Ni$^{2+}$ to be synthesized for the first time Fig.~\ref{fig:Periodic_Table}(a))~\cite{14_Cava_NaCaCo2F7,15_Krizan_NaCaNi2F7,16_Sanders_Na(Sr_Ca)(Mn_Fe)2F7}. 
However, this triumph comes at a cost: overall charge neutrality requires an effective 1.5+ valence state on the $A$-site, which necessitates the deliberate introduction of chemical disorder. 
The resulting chemical formula of the pyrochlore fluorides is $AA'B_2$F$_7$, where $A = $~Na$^+$ and $A'=$~Ca$^{2+}$, Sr$^{2+}$ are respectively alkali (Group 1) and alkaline earth (Group 2) metals in equal proportions, randomly distributed over the $A$ sublattice, as shown in Figure~\ref{fig:structures}(a) with crystallographic details provided in Table~\ref{table:crystal_structures}. 

Related fluoride pyrochlores in polycrystalline form were first reported in 1970~\cite{hansler1970a2b2f7}, but investigations of their magnetism have only recently begun following their growth as large single crystals using the floating zone technique~\cite{14_Cava_NaCaCo2F7}.
In contrast to the much studied rare earth pyrochlores, where the typical energy scale of the exchange interactions is 1~meV or smaller, the exchange interactions for magnets based on $3d$ transition metals are an order of magnitude or more larger. 
Therefore, whereas studying the correlated ground states of rare earth pyrochlores requires extremely low temperatures, well below 1~K, the correlated states of the fluoride pyrochlores can be studied at more easily accessible temperatures. 
All fluoride pyrochlores that have been characterized to date share two key attributes: (i) they are highly frustrated and (ii) that frustration is quenched at low temperatures by a spin freezing transition due to exchange disorder wrought by the intrinsic chemical disorder. 
The spin freezing transitions occur at a temperature, $T_f$ commensurate with the energy scale of the bond disorder, which is given by $\Delta/k_B = T_f \sqrt{\sfrac{3}{8}} \approx O(10^0)$~K~\cite{saunders2007spin}, while correlations onset at significantly higher temperatures, $|\theta_{\text{CW}}| \approx O(10^2)$~K. 
Thus, much of the experimental effort on fluoride pyrochlores has centered on characterizing their correlated states below their Curie-Weiss temperatures, $|\theta_{\text{CW}}|$, and above their spin freezing temperatures, $T_f$.

\subsection{Cobalt Fluoride Pyrochlores}

The cobalt-based fluoride pyrochlore, NaCaCo$_2$F$_7$, was first reported in 2014~\cite{14_Cava_NaCaCo2F7} and its sister compound, NaSrCo$_2$F$_7$, was subsequently reported in 2015~\cite{krizan2015nasrco2f7}. 
While more experimental studies have focused on the former compound, all available data suggests that the properties of these two compounds are largely indistinguishable. 
It is interesting to note that this strongly contrasts with rare earth oxide pyrochlores, where the choice of the non-magnetic $B$-site can dramatically modify the magnetic ground state~\cite{10_Gardner_Magnetic_pyrochlore_oxides}. 
Both NaCaCo$_2$F$_7$ ($\theta_{\text{CW}} = -140$~K, $T_f = 2.4$~K) and NaSrCo$_2$F$_7$ ($\theta_{\text{CW}} = -127$~K, $T_f = 3.0$~K) have strongly frustrated antiferromagnetic interactions, with frustration indices of 58 and 42, respectively~\cite{14_Cava_NaCaCo2F7,krizan2015nasrco2f7}. 
The spin freezing transitions in these materials, which are marked by a frequency dependent cusp in ac susceptibility, occur at a temperature commensurate with the strength of the chemical disorder. 
Thus, the slight increase in the freezing temperature of NaSrCo$_2$F$_7$ can be understood as originating from enhanced bond disorder due to the larger difference in ionic radii between Na ($r = 1.18$~\AA) and Sr ($r = 1.26$~\AA), as compared to Na and Ca ($r = 1.12$~\AA).

The single ion properties of the Co$^{2+}$ ions in these compounds are quite interesting in-and-of-themselves. 
This was initially hinted by a Curie-Weiss analysis of their magnetic susceptibility data, which revealed a paramagnetic moment around 6 $\mu_B$/Co$^{2+}$~\cite{14_Cava_NaCaCo2F7,krizan2015nasrco2f7}, dramatically enhanced from the spin-only value for high spin Co$^{2+}$ in a pseudo-octahedral environment ($S = \sfrac{3}{2}$, $\mu_{\text{calc}} = 3.87$~$\mu_B$). 
Instead, the observed moment is much closer to the value expected with a full orbital contribution ($J = \sfrac{9}{2}$, $\mu_{\text{calc}} = 6.63$~$\mu_B$). 
A subsequent inelastic neutron scattering study of the crystal field excitations in both NaCaCo$_2$F$_7$ and NaSrCo$_2$F$_7$ confirmed the intermediate spin-orbit coupling, which also leads to strong deviations from Heisenberg type moments~\cite{17_Ross_NaACo2F7}. 
Instead, the Co$^{2+}$ moments in these materials have relatively strong XY anisotropy, preferentially lying in the plane perpendicular to the local [111] direction, which is the axis that connects adjacent tetrahedra. 
This type of anisotropy is also found in the erbium and ytterbium families of rare earth oxide pyrochlores~\cite{18_Hallas_XY_pyrochlores}, as well as possibly the dysprosium chalcogenide spinels~\cite{13_wong_XY-pyrochlore}. 
An open question surrounds the complete absence of anisotropy in magnetization measurements for NaCaCo$_2$F$_7$~\cite{zeisner2019magnetic}, which does not conform with the expected behavior of an XY antiferromagnet~\cite{bonville2013magnetization}.

\begin{figure}
    \centering
    \includegraphics[width=3.2in]{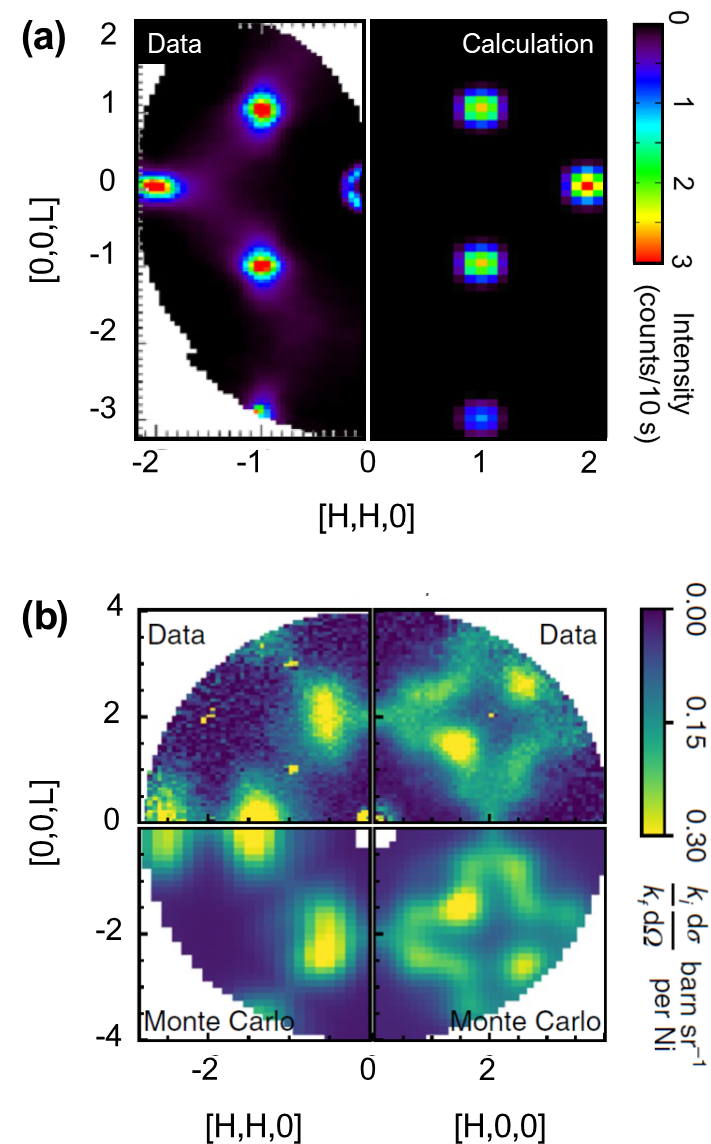}
    \caption{Elastic neutron scattering measurements of (a) NaCaCo$_2$F$_7$ and (b) NaCaNi$_2$F$_7$ below their spin freezing transitions, $T_f$, reproduced from Ref.~\cite{16_Ross_short_range_order_NaCaCo2F7} and~\cite{19_Plumb_NaCaNi2F7}, respectively. The elastic scattering of NaCaCo$_2$F$_7$ consists of broadened magnetic Bragg peaks and zigzag diffuse scattering. The former originates from short range antiferromagnetic XY order, and the latter originates from low energy excitations in the XY plane. The diffuse scattering for NaCaNi$_2$F$_7$ is well described by the Heisenberg Hamiltonian. Sharp pinch points, for example near (1 1 3), indicate there is zero average magnetization per tetrahedra.}
    \label{Fluorides_Neutron}
\end{figure}

Although the ultimate ground states of NaCaCo$_2$F$_7$ and NaSrCo$_2$F$_7$ are frozen glass-like states, they are not characterized by random spin configurations. 
Correlations with a typical length scale of a single tetrahedron set in at temperatures as high as 200~K and inter-tetrahedra correlations develop around 50 K, as observed by neutron scattering, NMR, and ESR measurements~\cite{frandsen2017real,sarkar2017spin,zeisner2019magnetic}. 
Elastic neutron scattering measurements below the spin freezing transition, shown in Fig.~\ref{Fluorides_Neutron}(a), show that the frozen state is made up of short-range ordered clusters with antiferromagnetic XY spin configurations ($\Gamma_5$ in the language of irreducible representations, shown in Fig.~\ref{fig:irreps} (b)) with a correlation length of 16~\AA~\cite{16_Ross_short_range_order_NaCaCo2F7}. 
Two unique, nearly degenerate ordered states exist within $\Gamma_5$, and the spin freezing transition preempts the selection of a unique ordered state in the cobalt pyrochlores. 
While the low energy spin excitations are consistent with a continuous degree of rotational symmetry in the local XY plane for these $\Gamma_5$ clusters, at higher energies a different type of spin excitation emerges. 
This higher energy spin excitation can be rather accurately captured by collinear antiferromagnetic correlations on individual tetrahedra, breaking the local XY anisotropy~\cite{16_Ross_short_range_order_NaCaCo2F7}. 
The presence of these two distinct excitations is suggestive of a frustration between intra-~and inter-tetrahedral correlations.

Another cobalt-based pyrochlore lattice material is realized in the carbonate Na$_3$Co(CO$_3$)$_2$Cl, which forms in the $Fd\bar{3}$ space group~\cite{zheng2011spin}. 
Unlike the fluorides, Na$_3$Co(CO$_3$)$_2$Cl is a fully site ordered material with no inherent disorder. 
The energy scale of this material's spin correlations, which is roughly parameterized by the Curie-Weiss temperature $\theta_{\text{CW}} = -34$~K~\cite{fu2013coexistence}, is reduced from the fluorides by nearly an order of magnitude. 
This is at least partially attributable to the larger Co-Co bond distance, 4.9~\AA in Na$_3$Co(CO$_3$)$_2$Cl as compared to 3.7~\AA~in the fluorides. 
This material has a particularly complex phase behavior, first entering a short-range correlated state at 17 K, exhibiting a spin-glass like transition at $T_f = 4.5$~K, before finally proceeding through an apparent long range ordering transition at $T_N = 1.5$~K~\cite{fu2013coexistence}. 
The ordered state is an all-in/all-out antiferromagnet, shown in Fig.~\ref{fig:irreps}~(a), which implies the Co$^{2+}$ moments in this material have Ising local anisotropy unlike the XY anisotropy observed in the fluorides. 
Future studies should confirm that this complex phase behavior is intrinsic and not the result of sample inhomogeneity or phase separation.

\subsection{Nickel Fluoride Pyrochlores}
  
The nickel fluoride pyrochlore NaCaNi$_2$F$_7$ ($\theta_{\text{CW}} = -129$~K, $T_f = 3.6$~K) is another highly frustrated antiferromagnet, with a frustration index of 36~\cite{15_Krizan_NaCaNi2F7}. 
The effective moment derived from Curie-Weiss fits, $3.7$~$\mu_B$, is larger than what is typically observed for the spin-only moment of Ni$^{2+}$ ($S = 1$, $\mu_{\text{calc}} = 2.83$~$\mu_B$) but falls significantly short of what would be expected for a full orbital contribution ($J = 1$, $\mu_{\text{calc}} = 5.59$~$\mu_B$)~\cite{15_Krizan_NaCaNi2F7}. 
Thus, spin-orbit coupling, while not negligible, is less significant than in the case of the cobalt analogs, placing NaCaNi$_2$F$_7$ closer to the Heisenberg limit. 
This picture is further validated by polarized neutron scattering measurements, where it was observed that the scattering in the spin flip and non-spin flip channels is nearly identical, indicating the interactions are highly isotropic~\cite{19_Plumb_NaCaNi2F7}.

While it is true that bond disorder induces a spin freezing in each of the fluoride pyrochlores, irrespective of the magnetic ion involved, their magnetic correlations do in fact differ substantially. 
In the case of the cobalt variants, spin freezing appears to preempt long-range magnetic order, while in the case of NaCaNi$_2$F$_7$, there is no indication that the compound is close to magnetic order. 
Elastic neutron scattering measurements in the frozen state, shown in Fig.~\ref{Fluorides_Neutron}(b), reveal structured diffuse scattering with a correlation length of 6~\AA~\cite{19_Plumb_NaCaNi2F7}, close to the next nearest neighbor bond distance. 
Fits to the neutron scattering data show that NaCaNi$_2$F$_7$ is very close to an exact realization of the nearest neighbor Heisenberg antiferromagnet with $S=1$, where exchange anisotropy, next nearest neighbor exchange, and bond disorder all represent small corrections~\cite{19_Plumb_NaCaNi2F7}. 
Sharp pinch point features, as famously observed in the dipolar spin ices, indicate that there is zero net magnetization on each tetrahedron (Fig.~\ref{Fluorides_Neutron}(b))~\cite{19_Plumb_NaCaNi2F7,zhang2019dynamical}. 
This observation in particular is suggestive of the classical Heisenberg pyrochlore spin liquid first studied by Villain~\cite{villain1979insulating}, where the ground state manifold is comprised of all states with zero magnetization per tetrahedron, yielding a macroscopic degeneracy. 
This may be consistent with the residual entropy of $0.176R$ per mol Ni observed in NaCaNi$_2$F$_7$, which is only 16\% of the expected entropy release for $R\ln{(3)}$~\cite{19_Plumb_NaCaNi2F7}.

Despite ultimately undergoing a spin freezing transition, quantum effects appear to be quite significant in NaCaNi$_2$F$_7$.
Neutron scattering measurements show that at low temperatures, 90\% of the magnetic scattering occurs inelastically, dramatically exceeding the semiclassical expectation of 50\% for $S=1$~\cite{19_Plumb_NaCaNi2F7}. 
In contrast, the cobalt variants agree well with this semiclassical picture where the observed 30\% elastic approaches the expected 33\% for $J=\sfrac{1}{2}$~\cite{16_Ross_short_range_order_NaCaCo2F7}. 
Further evidence for the importance of quantum fluctuations in NaCaNi$_2$F$_7$ comes from $\mu$SR measurements, which show the observed static field in the frozen state is an order of magnitude smaller than expected, perhaps signalling that the Ni moments are not fully frozen~\cite{18_Cai_MuSR_NaCaNi2F7}.

\subsection{Manganese, Iron, and Future Prospects} 

\begin{figure*}[htbp]
    \centering
    \includegraphics[width=7in]{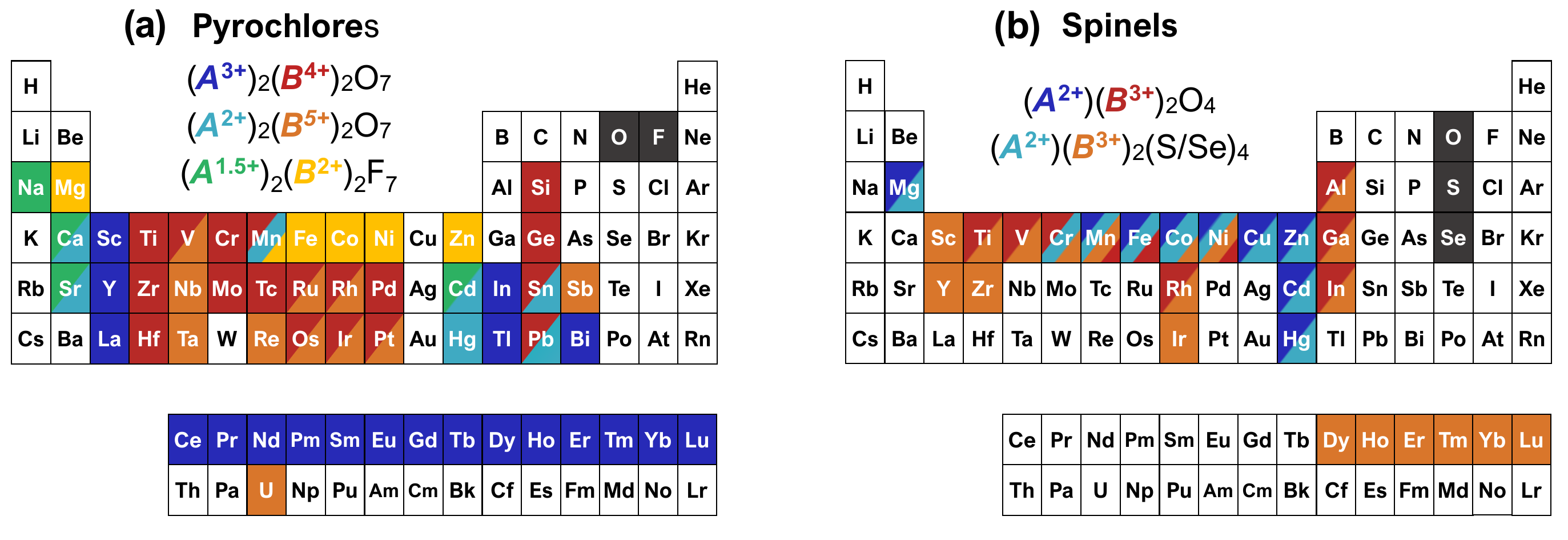}
    \caption{The various metals that are known to form (a) oxide and fluoride pyrochlores ($A_2^{3+}B_2^{4+}$O$_7$, $A_2^{2+}B_2^{5+}$O$_7$, and $A_2^{1.5+}B_2^{2+}$F$_7$) and (b) oxide and chalcogenide spinel ($A^{2+}B_2^{3+}$O$_4$ and $A^{2+}B_2^{3+}$(S/Se)$_4$). Of particular note, pyrochlores, in panel (a), with late $3d$ transition metals can only be formed as fluorides, and spinels, in panel (b), with rare earth ions can only be formed as sulfides or selenides.}
    \label{fig:Periodic_Table}
\end{figure*}

The fluoride pyrochlore family is rounded out by a trio of compounds, NaSrMn$_2$F$_7$ ($\theta_{\text{CW}} = -90$~K, $T_f = 2.5$~K), NaSrFe$_2$F$_7$ ($\theta_{\text{CW}} = -98$~K, $T_f = 3.7$~K), and NaCaFe$_2$F$_7$ ($\theta_{\text{CW}} = -73$~K, $T_f = 3.9$~K)~\cite{16_Sanders_Na(Sr_Ca)(Mn_Fe)2F7}. 
Following the familial trend, all three are highly frustrated antiferromagnets with frustration indices of 36, 26, and 19, respectively. 
In all three, the calculated paramagnetic moment is slightly larger than what would be expected for high spin $3d^5$ Mn$^{2+}$ ($S=\sfrac{5}{2}$) and high spin $3d^6$ Fe$^{2+}$ ($S=2$), respectively~\cite{16_Sanders_Na(Sr_Ca)(Mn_Fe)2F7}. 
Similar to Ni$^{2+}$ ($S=1$), these three compounds are all likely good realizations of Heisenberg spins on the pyrochlore lattice. 
However, due to their more classical nature (larger $S$ values) they will make for an interesting comparison with NaCaNi$_2$F$_7$ in terms of how well they conform to the classical spin liquid state expected for an antiferromagnetic Heisenberg pyrochlore~\cite{moessner1998properties}. 
To that end, detailed neutron spectroscopic studies on these Mn and Fe variants will be of the utmost interest. 
One might also expect that it may be possible to extend this series to include other transition metal ions; while the strong Jahn-Teller effect in copper renders it an unlikely candidate, chromium may be a promising avenue to explore.

Two related Mn and Fe based compounds that bear mentioning here are Na$_3$Mn(CO$_3$)$_2$Cl and FeF$_3$, based on Mn$^{2+}$ and Fe$^{3+}$, respectively. 
These two ions have the same electronic configuration, $3d^5$ ($S=\sfrac{5}{2})$, and both compounds are good realizations of the classical Heisenberg pyrochlore antiferromagnet. 
Na$_3$Mn(CO$_3$)$_2$Cl shares the same crystal structure as the aforementioned Na$_3$Co(CO$_3$)$_2$Cl and, unlike the fluorides, does not possess intrinsic disorder. 
It does not magnetically order or undergo a spin freezing transition down to at least 0.5~K, which when coupled with a Curie-Weiss temperature of $-41$~K yields a large frustration index of at least $f = 80$~\cite{nawa2018degenerate}. 
Below 1~K, Na$_3$Mn(CO$_3$)$_2$Cl exhibits a rather sharp increase in its heat capacity that continues down to the lowest measured temperatures, 0.5~K, and the nature of this feature remains unresolved~\cite{nawa2018degenerate}.

Finally, FeF$_3$ with space group $Fd\bar{3}m$ is a unique material with no isomorphic compounds. 
The pyrochlore lattice of Fe$^{3+}$ is built from a network of corner sharing FeF$_6$ octahedra~\cite{de1986new}. 
Compared to other materials discussed in this section, all of which are relatively recently characterized, magnetic studies of FeF$_3$ date back to 1986~\cite{de1986new,ferey1986ordered}. 
This material ultimately magnetically orders close to 20~K in an antiferromagnetic all-in/all-out structure (Fig.~\ref{fig:irreps}~(a))\cite{ferey1986ordered,calage1987mossbauer}. 
However, intense frustration is evidenced by a lack of Curie-Weiss behavior in susceptibility up to at least 300~K and structured magnetic diffuse scattering up to at least 100~K~\cite{ferey1986ordered,reimers1991short}. 
Further studies of FeF$_3$ and Na$_3$Mn(CO$_3$)$_2$Cl would be facilitated by their growth in large single crystal form, which represents a significant challenge in both cases.

\section{Rare earth chalcogenide spinels} \label{RE-spinels}

Much like structural tolerance factors forbid the formation of oxide pyrochlores with late $3d$ transition metals, the large ionic radius of the $4f$ rare earth elements does not allow them to be accommodated by the oxide spinel structure.
Given the diversity of magnetic states that can be unlocked in the rare earth $R_2B_2$O$_7$ pyrochlores simply through replacement of the non-magnetic $B$-site~\cite{greedan2001geometrically}, the more significant shift to the spinel structure is an important prospect to search for new and exotic magnetic states. 
It is only through the replacement of oxygen by a larger chalcogen, sulfur or selenium, that rare earth elements can be incorporated into the spinel structure (Fig.~\ref{fig:Periodic_Table}(b)) -- and even then, it is only the heaviest rare earths, which have the smallest ionic radii due to the lanthanide contraction. 
In this family, compounds of the form $AR_2X_4$, with $X = \mathrm{S}$ and Se, are stable for $A =$~Cd and $R =$~Dy $-$ Yb, \cite{64_Suchow_fluores_RE_spinel} as well as for the slightly larger $A =$~Mg, with $R =$~Ho $-$ Yb \cite{65_Flahaut}, first reported in 1964 and 1965, respectively. Remarkably, replacing the $A$ site by any larger alkaline earth metal (Ca, Sr, or Ba) gives an assortment of orthorhombic and rhombohedral structure types~\cite{65_Flahaut}.

Many of the distinctive properties of the rare earth chalcogenide spinels are directly attributable to the modified local environment.
In this structure, with crystallographic details provided in Table~\ref{table:crystal_structures}, the rare earth sits in a distorted octahedral environment (Fig.~\ref{fig:structures}(b)) as opposed to the pseudo-cubic coordination of the rare earth site in oxide pyrochlores. 
This leads, in many cases, to an inversion of the single ion anisotropy (\emph{e.g.} ions that are Ising-like in the oxide pyrochlores are XY-like in the chalcogenide spinels). 
Furthermore, the coordinating anions are between 10 and 30\% farther away than in the pyrochlore oxides and chalcogens are themselves less electronegative than oxygen. 
Both of these factors serve to lower the energy scale of the crystal field splitting and this leads to more complex ground state manifolds than the well-isolated ground state doublets generally found in the rare earth pyrochlore oxides. 
Taken altogether, the chalcogenide spinels exhibit entirely different crystal electric field (CEF) ground states than their oxide pyrochlore counterparts, as highlighted with erbium in Fig.~\ref{fig:spinel_CEF}.

The crystal field states of the chalcogenide spinels were first studied in the 1970's~\cite{74_Pokrzywnicki_CEF_CdYb2Se4,75_Pokrzywnicki_CEF_CdYb2Se4,80_Ben-Dor_CdRE2S_series,88_pawlak_Mag_CEF_Yb,77_pokrzywnicki_mag_tm2se3_CdTm2Se4}. 
The accuracy of these studies was limited by their assumption of a perfect octahedral environment and also because magnetization data alone was used for the refinements of the CEF Hamiltonian. 
More recent studies have used inelastic neutron scattering to directly measure the CEF levels and have found that although the ligands form a near perfect octahedra, the field from the next nearest neighbor creates a significant anisotropy along the local [111] directions~\cite{18_Gao_CdEr2Se4, 18_Reig_MgEr2Se4,19_Reig_MgRE2Se4}. 
The other major omission of the works prior to the new millennia is that they did not take into account the importance of geometric frustration, which explains their findings that none of these materials magnetically order above 2 K \cite{72_Fuji_Mag_CdRE2X4,80_Ben-Dor_CdRE2S_series,74_Pokrzywnicki_CEF_CdYb2Se4,75_Pokrzywnicki_CEF_CdYb2Se4}. 
The importance of geometric frustration in these chalcogenide spinels was not considered until 2005~\cite{05_Lau_CdRE2X4}. 
The combination of strong anisotropy, complex crystal field ground states, and intense geometric magnetic frustration yields a collection of remarkable magnetic states in the rare earth chalcogenide spinels.

\subsection{Erbium Chalcogenide Spinels}

\begin{figure}[tb]
    \centering
    \includegraphics[width=3.2in]{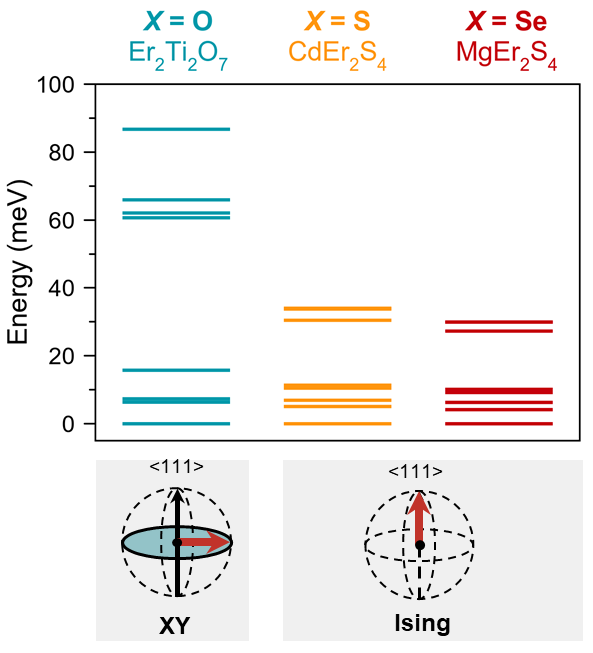}
    \caption{The crystal field splitting of Er$^{3+}$ in the pyrochlore oxide Er$_2$Ti$_2$O$_7$~\cite{gaudet2018effect} compared to the chalcogenide spinels CdEr$_2$S$_4$~\cite{18_Gao_CdEr2Se4} and MgEr$_2$Se$_4$~\cite{18_Reig_MgEr2Se4}. Due to the reduction in anion electronegativity, the overall magnitude of the crystal field splitting decreases significantly going from $X=$~O to $X=$~S and Se. The change in the local environment also produces an inversion of the anisotropy: Er${3+}$ in the pyrochlore oxide has XY anisotropy while in the chalcogenide spinels it has Ising anisotropy (perpendicular and parallel to the local [111], respectively).}
    \label{fig:spinel_CEF}
\end{figure}

In the erbium chalcogenide spinels, $A$Er$_2X_4$ ($A =$~Cd or Mg and $X=$~S or Se), the pyrochlore sublattice is occupied by Er$^{3+}$, which is a Kramers ion with a total angular momentum of $J = \sfrac{15}{2}$ and an expected paramagnetic moment of $\mu_{\text{calc}}=9.6$~$\mu_B$, which agrees well with experiment~\cite{05_Lau_CdRE2X4}. 
While erbium-based oxide pyrochlores have local XY anisotropy of varying strength~\cite{gaudet2018effect}, the change in the local environment in the spinels yields perfectly Ising moments~\cite{lago2010cder,18_Gao_CdEr2Se4,18_Reig_MgEr2Se4}, constrained to lie along the local [111] directions (Fig.~\ref{fig:spinel_CEF}). 
However, the first excited CEF levels are found at energies as low as 4~meV above the ground state~\cite{18_Reig_MgEr2Se4,18_Gao_CdEr2Se4} and as a result, their influence cannot be entirely disregarded. 
In particular, coupling between the ground and excited states through so-called virtual CEF fluctuations becomes appreciable at these energy scales~\cite{16_rau_virtual_CEF}, which can disrupt the otherwise perfect Ising anisotropy. 
Studies thus far suggest that the $A$Er$_2X_4$ compounds, with varying $A$ and $X$ ions, have broadly similar properties, which are to first order dictated by the strong Ising anisotropy.

The earliest studies on the erbium spinels predicted that these compound would order between 4 and 10 K, based on the the onset of short range order seen in magnetization and M\"ossbauer spectroscopy measurements~\cite{72_Fuji_Mag_CdRE2X4,81_ben-dor_CdEr2Se4_mag_and_mossbauer}. 
However, it wasn't until 2010 that the nature of the short range order was understood~\cite{lago2010cder}. 
The very large effective moments in the erbium spinels combined with their Ising anisotropy, gives rise to long range dipolar interactions that produce a ferromagnetic coupling~\cite{den2000dipolar}, significantly larger in magnitude than the antiferromagnetic near neighbour exchange~\cite{18_Reig_MgEr2Se4,18_Gao_CdEr2Se4}.
Thus, the erbium spinels possess all the requisite ingredients for spin ice physics and that is indeed what has been found.
This was first uncovered in CdEr$_2$Se$_4$, where heat capacity measurements showed the absence of long-range magnetic order but the presence of a broad heat capacity anomaly centered below 1~K~\cite{lago2010cder}. 
The remnant entropy associated with this heat capacity anomaly approaches the Pauling value, associated with the macroscopic degeneracy of ice-rules configurations~\cite{pauling1935,ramirez1999zero}.
The spin ice state of CdEr$_2$Se$_4$ ultimately drops out of thermal equilibrium, leading to a spin freezing transition at $T_f = 0.6$~K~\cite{lago2010cder}. 
A M\"ossbauer study on CdEr$_2$S$_4$ showed that the spin fluctuations are significantly slower than in the erbium oxide pyrochlores~\cite{15_Legros_CdEr2Se4_mossbauer}, a likely consequence of the the larger energy cost to flip an Ising moment.

The spin ice states observed in the erbium spinels have intriguing differences with the well known holmium and dysprosium oxide pyrochlores, in which the spin ice state was first discovered~\cite{harris1995magnetic}. 
A careful analysis of the heat capacity of MgEr$_2$Se$_4$, as well as the previously published data in CdEr$_2$Se$_4$, shows that, in fact, the remnant entropy is about 20\% lower than the expected Pauling entropy of a spin ice~\cite{18_Reig_MgEr2Se4}. 
In the same compound, Monte Carlo simulations with the dipolar spin ice model capture the qualitative features of the heat capacity and the magnetic diffuse scattering~\cite{18_Gao_CdEr2Se4}, but fail to quantitatively describe the data~\cite{18_Reig_MgEr2Se4}. 
AC susceptibility studies suggest that the monopole quasiparticle excitations in the ground state of CdEr$_2$Se$_4$ have three orders of magnitude higher mobility than in the prototypical spin ice Dy$_2$Ti$_2$O$_7$~\cite{18_Gao_CdEr2Se4}. 
These results suggest that there are additional interactions beyond the normal dipolar ice model, allowing both faster monopole dynamics and reducing the number of states in the degenerate ground state manifold. 
It is likely that these additional interactions stem from the much smaller energy gap between ground state and first exited CEF levels~\cite{18_Gao_CdEr2Se4,18_Reig_MgEr2Se4}.

\subsection{Ytterbium Chalcogenide Spinels}

The ytterbium spinels, $A$Yb$_2X_4$ ($A =$~Cd or Mg and $X=$~S or Se), have a pyrochlore network of Yb$^{3+}$, which has a Hund's rules total angular moment of $J = \sfrac{7}{2}$ and a paramagnetic moment of $\mu_{\text{calc}}=4.5$~$\mu_B$. 
In contrast to their erbium counterparts, the ytterbium spinels do have a well-isolated CEF ground state, a Kramers pseudo-spin-$\sfrac{1}{2}$ doublet which is separated by approximately 25~meV from the first excited state. 
There are conflicting reports on the spin anisotropy of the Yb$^{3+}$ moments, which can only partially be explained by material dependent parameters (the specific $A$ and $X$ ions). 
While one study found that the in- and out-of-plane components are nearly equal~\cite{17_Higo_AYb2X4} and thus, Heisenberg-like, others have reported that it is weakly~\cite{guratinder2019multiphase}, or moderately Ising-like~\cite{19_Reig_MgRE2Se4}. 
An exact identification of the ground state anisotropy is difficult because direct measurements of the CEF energy levels via neutron scattering leads to an underconstrained problem. 
An additional constraint can be added by considering the magnetization as a function of applied field~\cite{19_Reig_MgRE2Se4}, which led to the most Ising-like result. 
However the accuracy of this method is debatable, as the strength of short range correlations at low temperature may confound the fits~\cite{guratinder2019multiphase}. 
Electron paramagnetic resonance spectroscopy measurements on magnetically dilute samples could help resolve this question.

\begin{figure*}
\includegraphics[width=\linewidth]{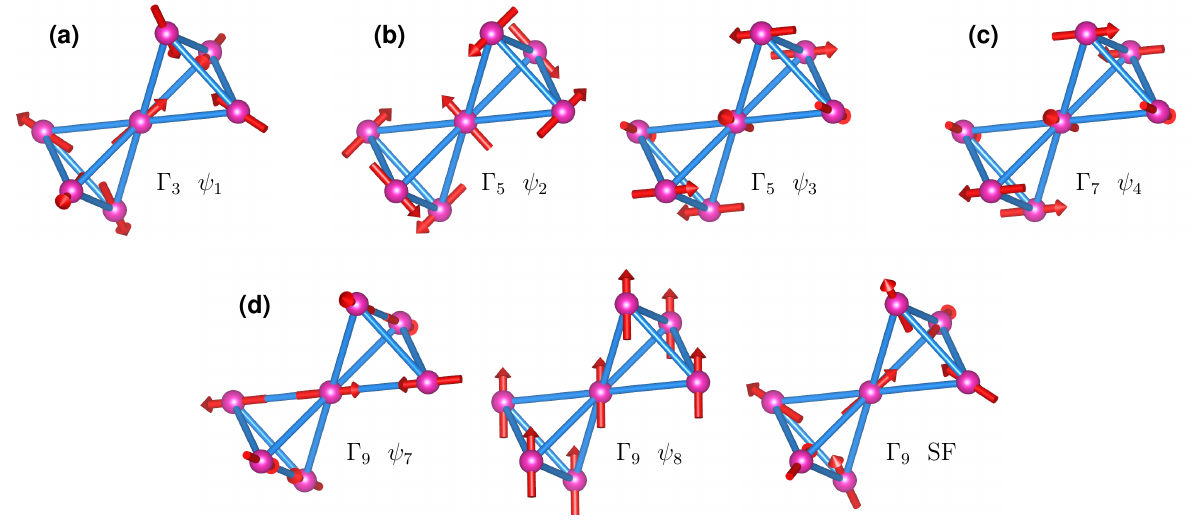}
\caption{The irreducible representations ($\Gamma$) and their basis vectors ($\psi$), for allowed $k=0$ magnetic orders on the pyrochlore lattice. (a) $\Gamma_3$ is commonly known as the all-in all-out structure. (b) $\Gamma_5$ is usually found in compounds with XY anisotropy and is composed of linear combinations of $\psi_2$ and $\psi_3$. (c) $\Gamma_7$, often referred to as the Palmer-Chalker state, has a threefold basis, where $\psi_5$ and $\psi_6$ are equivalent to $\psi_4$ rotated about the primary crystal axis. (d) $\Gamma_9$ has a sixfold basis, where we show two representative basis vectors, $\psi_7$ and $\psi_8$. The other four basis states ($\psi_9$, $\psi_{10}$, $\psi_{11}$, and $\psi_{12}$) are related to $\psi_7$ and $\psi_8$ by a simple rotation to an equivalent cubic direction. The linear combinations of $\psi_7$ and $\psi_8$ forms the ice-like splayed ferromagnet state (SF) shown. Other linear combinations can give ferromagnetic states with the moment canted away from the local [111].}
\label{fig:irreps}
\end{figure*}

All four ytterbium spinels have long-range antiferromagnetic ordering transitions between 1 and 2~K~\cite{17_Higo_AYb2X4}. Neutron diffraction has only been performed in the two $A$ = Cd compounds, both of which order into the $\Gamma_5$ irreducible representation~\cite{17_de_reotier_CdYb2X4,guratinder2019multiphase}, which is pictured in Fig~\ref{fig:irreps}(b), where linear combinations of the basis vectors $\psi_2$ and $\psi_3$ trace out the local XY plane. 
The ordered moment is significantly reduced compared to the calculated value for the CEF ground state doublet, indicating that strong quantum fluctuations are at play~\cite{17_de_reotier_CdYb2X4,guratinder2019multiphase}. 
Spontaneous muon precession is observed in CdYb$_2X_4$~\cite{17_de_reotier_CdYb2X4} but not in MgYb$_2$S$_4$~\cite{17_Higo_AYb2X4}, suggesting that a different ground state might be found in the Mg analogs, or they may have stronger disorder. 
Within the ordered state, gapless spin excitations are observed in both electron spin resonance measurements~\cite{Yoshizawa2015} and heat capacity~\cite{17_Higo_AYb2X4}. 
It is interesting to note that there are several order-by-disorder mechanisms through which the $\Gamma_5$ degeneracy can be lifted, which would open a small energy gap that as of now has not been detected~\cite{18_Rau_goldsone_in_order_by_disorder}.
One of these mechanisms, virtual CEF fluctuations, can likely be excluded, as the splitting to the first excited CEF level is rather large. 
Muon spin relaxation studies show that there are persistent spin dynamics well below the ordering temperature~\cite{17_de_reotier_CdYb2X4}, a feature that is apparently ubiquitous in rare earth pyrochlore lattice materials, irrespective of anion~\cite{mcclarty2011calculation,15_Yaouanc_CdHo2S4,16_deReotier_muSR_pyrochlores}.

Although the moments of the ytterbium spinels are not in and of themselves XY-like, there is tantalizing evidence to suggest shared phenomenology with the XY oxide pyrochlores, whose properties are dictated by intense phase competition. 
The first indication comes from heat capacity, where it is observed that the sharp lambda-like ordering anomalies are preceded in temperature by a broad anomaly, such that only 30\% of the entropy release occurs below $T_N$~\cite{17_de_reotier_CdYb2X4,17_Higo_AYb2X4}. 
This two-stage ordering has become synonymous with phase competition~\cite{18_Hallas_XY_pyrochlores}. 
More direct evidence of competing phases comes from the low magnetic fields required to destabilize their magnetic ground states, as was first observed in CdYb$_2$S$_4$ with electron spin resonance~\cite{Yoshizawa2015}. 
At moderate fields there is a crossover from the XY antiferromagnetic $\Gamma_5$ state (Fig.~\ref{fig:irreps}~(b)) to a spin-ice like ferromagnetic $\Gamma_9$ state (Fig.~\ref{fig:irreps}~(d))~\cite{guratinder2019multiphase}. 
Theoretical modeling further reinforces the possible importance of phase competition in this family of compounds, showing that CdYb$_2$Se$_4$ lies at a classical phase boundary separating the splayed ferromagnetic and antiferromagnetic $\Gamma_5$ phase~\cite{18_Rau_Yb_pyrochlore}. 
However, inelastic neutron scattering reveals weakly dispersive spin excitations within its ordered state that do not, at face value, resemble the overdamped spin excitations found in the ytterbium oxide pyrochlore family~\cite{hallas2016universal}.
Future works should clarify the role of phase competition and order-by-disorder effects in the Yb spinels, for which single crystal samples will be crucial.

\subsection{Holmium Chalcogenide Spinels}

The holmium chalcogenide spinels are still poorly understood due to their complex single ion properties. 
The Ho$^{3+}$ cation that occupies the pyrochlore lattice has a total angular momentum of $J=8$ ($\mu_{\text{calc}} = 10.6$~$\mu_B$) and is hence a non-Kramers ion where a ground state doublet is no longer guaranteed. 
The CEF ground state was initially suggested to be a non-magnetic singlet~\cite{80_Ben-Dor_CdRE2S_series}, due to low temperature susceptibility measurements showing temperature independent paramagnetism in CdHo$_2$S$_4$. 
More recently, direct measurement of the CEF levels in MgHo$_2$Se$_4$ via inelastic neutron scattering reveals a non-Kramers doublet ground state, in close proximity to an excited doublet and singlet below 1~meV, and yet another singlet below 3~meV~\cite{19_Reig_MgRE2Se4}.
This complicated CEF scheme with many low lying levels bares some resemblance to the terbium oxide pyrochlores, whose ground states are formed by two closely spaced doublets. In the terbium pyrochlores, the competing energy scales of single ion effects and the ion-ion interactions leads to rich and complex physics that has eluded description by a microscopic theory~\cite{07_molavian_QSI_tb_pyrochlore,12_Petit_Tb,19_Rau_Frustrated_quantum_RE_pyrochlores,hallas2020intertwined}. 
In the holmium spinels the lowest lying excited CEF levels are even more numerous and closely spaced than the 1.5~meV level in the terbium pyrochlores, and thus we can expect similarly rich physics.

The magnetic ground state of the holmium chalcogenides has only been studied in detail in one compound, CdHo$_2$Se$_4$, which has a magnetic transition at $T_C= 0.87$~K~\cite{15_Yaouanc_CdHo2S4}. 
However, as is the case for the terbium pyrochlores, this ordering transition appears unconventional; despite the very sharp anomaly in heat capacity, magnetic susceptibility only shows a broad inflection at $T_C$. 
Muon spin relaxation measurements indicate significant fluctuations in the ordered state that persist in the paramagnetic regime above $T_C$ with a fluctuation rate that is two orders of magnitude slower than expected~\cite{15_Yaouanc_CdHo2S4}.
Given the non-Kramers nature of Ho$^{3+}$ and its complex CEF ground state, this effect may be related to a muon-induced local distortion rather than an intrinsic origin~\cite{foronda2015anisotropic}. 
Neutron scattering studies of the static and dynamic properties of the holmium chalcogenide spinels below $T_C$ will be illuminating.

\subsection{Dysprosium, Thulium, and Future Prospects}

There are additional members of the rare earth chalcogenide spinel family with either thulium or dysprosium on the pyrochlore sublattice, for which experimental measurements are still in their infancy. 
However, what is known coupled with theoretical predictions suggests rich physics that warrants further research.

MgTm$_2$Se$_4$ has a spin singlet CEF ground state with a second singlet level within 1~meV~\cite{19_Reig_MgRE2Se4}. 
This pair of narrowly separated singlets is analogous to a regular Kramers doublet that is split by an external magnetic field, equivalent to an Ising system split by a transverse field~\cite{20_liu_TFIM}. 
This is a particularly interesting case because, for the local symmetry of the pyrochlore lattice (where the Ising moments point along eight equivalent local [111] directions and hence are not colinear), such a transverse field would be impossible to reproduce with the application of an external field. 
The exact ground state would then depend on the relative strength of the singlet-singlet CEF level spacing and the nearest neighbor exchange, with the most interesting scenario being where they are of a comparable magnitude, possibly leading to a spin liquid state~\cite{20_liu_TFIM}.
While experimental studies of these compounds are limited, we know that there is no magnetic order in CdTm$_2$S$_4$ down to 2 K~\cite{05_Lau_CdRE2X4}, or in MgTm$_2$Se$_4$ down to 0.4 K~\cite{mauthesis}.

The CEF ground state of the dysprosium spinels has not been directly determined. 
However, a scaling analysis of the CEF parameters for the isostructural Er compound suggest that CdDy$_2$Se$_4$ would have XY spin anisotropy, although with a relatively small CEF separation of only 1.6 meV~\cite{13_wong_XY-pyrochlore}.
Even disregarding the effects of the low-lying excited CEF levels, CdDy$_2$Se$_4$ has been proposed as a candidate for quantum order-by-disorder or a U(1) quantum spin liquid~\cite{13_wong_XY-pyrochlore}. 
If the exchange interaction in CdDy$_2$Se$_4$ favor ordering along the local [1 1 1] directions, then the expected large in-plane component of the spin anisotropy would allow for strong quantum fluctuations and could be a candidate for a quantum spin ice state~\cite{15_Petrova_dilute_QSI}.

Chalcogenide spinels with rare earths larger than dysprosium have not been shown to be stable under regular synthesis conditions. 
However preparation of compounds with other rare earth elements in the $B$ site may be possible under high pressure conditions. 
High pressure methods have been successfully applied to expanding the family of oxide pyrochlores~\cite{wiebe2015frustration} but have not yet, to our knowledge, been tested with other anions.

Although we focused here on compounds where only the pyrochlore sublattice is occupied by a magnetic ion, there exist chalcogenide rare earth spinels with either Mn or Fe on the $A$-site. 
These materials have either surprisingly low~\cite{92_Pawlak_TMYb2Se4}, or no observed~\cite{67_Longo_MnYb2S4,70_Suchow_suscept_RE_selenide_spinels,92_Pawlak_TMYb2Se4,71_Kainuma_mag_prop_RE_spinels} magnetic ordering transitions down to liquid helium temperatures, pointing to the importance of geometric frustration. 
In fact, of the chalcogenides spinels with magnetic ions on both sites, it is only those with significant site mixing and disorder that have been observed to magnetically order~\cite{80_Tomas_FeYb2S4,93_Pawlak_TMSc2S4,17_Tsurkan_Fe_off_stoich}. 
The nature of the magnetic frustration in this family is still poorly understood, and the inclusion of a magnetic atom on the $A$-site adds an entirely new layer of complexity that is beyond the scope of pyrochlore-lattice frustration.

Given the highly anisotropic nature of rare earth magnetism, many of the most illuminating experiments, including inelastic neutron scattering, are best accomplished with single crystal samples. Thus far, the rare earth chalcogenide spinels have only been produced in polycrystalline form.
Flux crystal growth is a promising avenue as related chalcogenide compounds have been successfully grown with arsenic and antimony chalcogenide fluxes~\cite{74_Bohac_flux_methods_chalcogines}, or with cadmium chloride flux~\cite{71_von_Phillipsborn_crystal_growth}.
A second possible route is vapor transport methods. Similar cadmium chalcogenide spinels have been successfully grown with iodine or cadmium chloride as a transport agent~\cite{71_von_Phillipsborn_crystal_growth}.
Studies to find suitable single crystal growth methods are demanding, but such samples would open up many new avenues of research.

\section{Chromium Chalcogenide Spinels} \label{Cr-spinel}

In contrast to the other materials discussed in this review, the family of chromium spinels $A$Cr$_2X_4$ ($A=$~Cd, Co, Cu, Fe, Mn, Zn; $X=$~S, Se), where chromium occupies the pyrochlore sublattice, have been extensively studied dating back well over 50 years~\cite{71_von_Phillipsborn_crystal_growth}. 
The chromium spinels are unique in that, despite being a family of compounds with only one  magnetic ion, they span a wide range of magnetic phenomena. 
The Cr$^{3+}$ ion is an excellent example of an isotropic $S = 3/2$ moment with several competing exchange interactions that are highly dependent on lattice spacing~\cite{07_Rudolf_Cr_spinel}. 
This gives a near continuum of dominant interactions, from the geometrically frustrated antiferromagnetic chromium oxides such as ZnCr$_2$O$_4$ ($\theta_{\text{CW}} = -398$~K~\cite{02_Lee_ZnCr2O4}), to the bond frustrated ZnCr$_2$S$_4$ ($\theta_{\text{CW}} = 7.9$~K~\cite{07_rudolfZnCr2S4}), to the ferromagnetic HgCr$_2$Se$_4$ ($\theta_{\text{CW}} = 200$~K~\cite{baltzer1966exchange}). Within this large range of interaction strengths, we find spin-spiral~\cite{06_Chern}, phase coexistence~\cite{Hamedoun_1986}, and compounds with ferromagnetic spin fluctuations preceding antiferromagnetic order~\cite{78_akimitsu}.
Beyond their magnetic ground states, the physics of chromium spinels extend far beyond the scope of this review including semiconductors~\cite{baltzer1966exchange}, multiferroics~\cite{05_Hemberger_CdCr2S4_ferroelectric}, Chern semimetals~\cite{xu2011chern}, and colossal magnetoresistance~\cite{ramirez1997colossal}. We will not attempt to be complete on this topic and will instead refer the reader to other comprehensive reviews~\cite{thota2017magnetic}.

\begin{table*}[htbp]
\caption{ An overview of the magnetic properties of the materials discussed in this review. For each family and magnetic ion, one representative compound is chosen. Details include the angular momentum ($S$ for transition metals, and $J$ for rare earths), the expected moment size ($\mu_{calc}$), the paramagnetic moment determined from Curie-Weiss fitting ($\mu_{\text{eff}}$), the single-ion anisotropy, the Curie-Weiss temperature ($\theta_{CW}$), transition temperatures, and a brief description of the magnetic ground state.}
\begin{ruledtabular}
\begin{tabular}{l c c c c c c c }
\multicolumn{8}{c}{\textbf{Fluoride Pyrochlores and Related Materials}}\\ 
\hline 
   & $S$ & $\mu_{\text{calc}}^a$ ($\mu_\text{B}$) & $\mu_{\text{eff}}$ ($\mu_\text{B}$) & Anisotropy & $\theta_{\text{CW}}$ (K) & Transition temp. (K) &   Ground state \\
\hline
NaCaNi$_2$F$_7$ \cite{15_Krizan_NaCaNi2F7} & 1 & 2.83  & 3.7 & Heisenberg & $-129$ & T$_f$ = 3.6 &  Spin glass \\
NaCaCo$_2$F$_7$ \cite{14_Cava_NaCaCo2F7} & 3/2 & 3.87 & 6.1 & XY \cite{17_Ross_NaACo2F7} & $-139$ & T$_f$ = 2.4 &  Spin glass  \\
Na$_3$Co(CO$_3$)$_2$Cl \cite{fu2013coexistence}& 3/2 & 3.87 & 5.3 & Ising$^b$ & $-34$ & T$_N$ = 1.5 &  AFM$^c$ (All-in/all-out) \\
NaSrFe$_2$F$_7$ \cite{16_Sanders_Na(Sr_Ca)(Mn_Fe)2F7} & 2 & 4.90 & 5.94 & Heisenberg & $-98$ & T$_f$ = 3.7 & Spin glass \\
NaSrMn$_2$F$_7$ \cite{16_Sanders_Na(Sr_Ca)(Mn_Fe)2F7} & 5/2 & 5.92 & 6.25 & Heisenberg & $-90$ & T$_f$ = 2.5 &  Spin glass \\
Na$_3$Mn(CO$_3$)$_2$Cl \cite{nawa2018degenerate} & 5/2 & 5.92 & 5.97 & Heisenberg & $-41$ &  & No order to 0.5 K  \\
FeF$_3$ \cite{calage1987mossbauer}  & 5/2 & 5.92 & n/a$^d$ & Ising$^b$ & n/a$^d$ & T$_N$ = 21.8 &   AFM (All-in/all-out) \\

& & & & & & &\\

\multicolumn{8}{c}{\textbf{Rare Earth Chalcogenide Spinels}}\\
\hline
 & $J$ & $\mu_{\text{calc}}$ ($\mu_\text{B}$) & $\mu_{\text{eff}}$ ($\mu_\text{B}$) & Anisotropy & $\theta_{\text{CW}}$ (K) & Transition temp. (K) &   Ground state \\
\hline

CdDy$_2$Se$_4$ \cite{05_Lau_CdRE2X4} & 15/2 & 10.65 & 10.76  & XY \cite{13_wong_XY-pyrochlore} & $-7.6$ &  & Unknown \\
CdHo$_2$S$_4$ \cite{15_Yaouanc_CdHo2S4} & 8 & 10.8 & 10.8 & See text & $-3.6$ & T$_N = 0.9$ & Unknown \\ 
CdEr$_2$Se$_4$ \cite{lago2010cder} & 15/2 & 9.58 & 9.6 \cite{05_Lau_CdRE2X4} & Ising & $-1.2$  & T$_f = 0.6$ & Spin ice \\
CdTm$_2$S$_4$  \cite{05_Lau_CdRE2X4} & 6 & 7.56 & 7.58& Singlet \cite{19_Reig_MgRE2Se4} & $-11.8$ &  & Spin singlet \\
CdYb$_2$S$_4$ \cite{17_Higo_AYb2X4}  & 7/2 & 4.53 & 4.41 \cite{05_Lau_CdRE2X4} & See text &  $-10$  & $\mathrm{T}_N = 1.8$ & AFM ($\Gamma_5$ order)  \\

& & & & & & &\\

\multicolumn{8}{c}{\textbf{Breathing Pyrochlore Chalcogenide Spinels}}\\ 

\hline
  & $S$ & $\mu_{\text{calc}}^a$ ($\mu_\text{B}$) & $\mu_{\text{eff}}$ ($\mu_\text{B}$) & Anisotropy & $\theta_{\text{CW}}$ (K) & Transition temp. (K) &   Ground state \\
\hline

LiGaCr$_4$S$_8$ \cite{pokharel2018negative} & 3/2 & 3.86 & 3.96 & Heisenberg & 19.5 & T$_f = 10$ &   Cluster glass \\
CuInCr$_4$Se$_8$ \cite{plumier1971magnetic} & 3/2 & 3.86 & 3.58 & Heisenberg & 100 &  & Sample dep. \\
CuInCr$_4$S$_8$ \cite{plumier1971magnetic} & 3/2 & 3.86 & 3.83 & Heisenberg & $-77$ & T$_N = 40$ & AFM \\
CuGaCr$_4$S$_8$ \cite{wilkinson1976magnetic} & 3/2 & 3.86 &  & Heisenberg & & T$_N = 4.2$ & AFM (Incomm. spiral) \\
CuGaCr$_4$S$_8$ \cite{plumier1971mise} & 3/2 & 3.86 &  & Heisenberg & 142 & T$_N = 17$ & AFM (Incomm. spiral) \\

\end{tabular}
\end{ruledtabular}

\vspace{1ex}

 {\raggedright $^a$ For transition metal compounds, the calculated moment is the spin only contribution. $^b$ Ising spins are implied by the all-in/all-out ordered ground state of these compounds. $^c$ Antiferromagnet (AFM). $^d$ FeF$_3$ does not exhibit Curie-Weiss like susceptibility up to the highest measured temperatures. \par}

\label{table:compound_summary}
\end{table*}

\subsection{Breathing pyrochlore chromium spinels} \label{breathing-pyrochlore}

\begin{figure}
    \centering
    \includegraphics[width=3.5in]{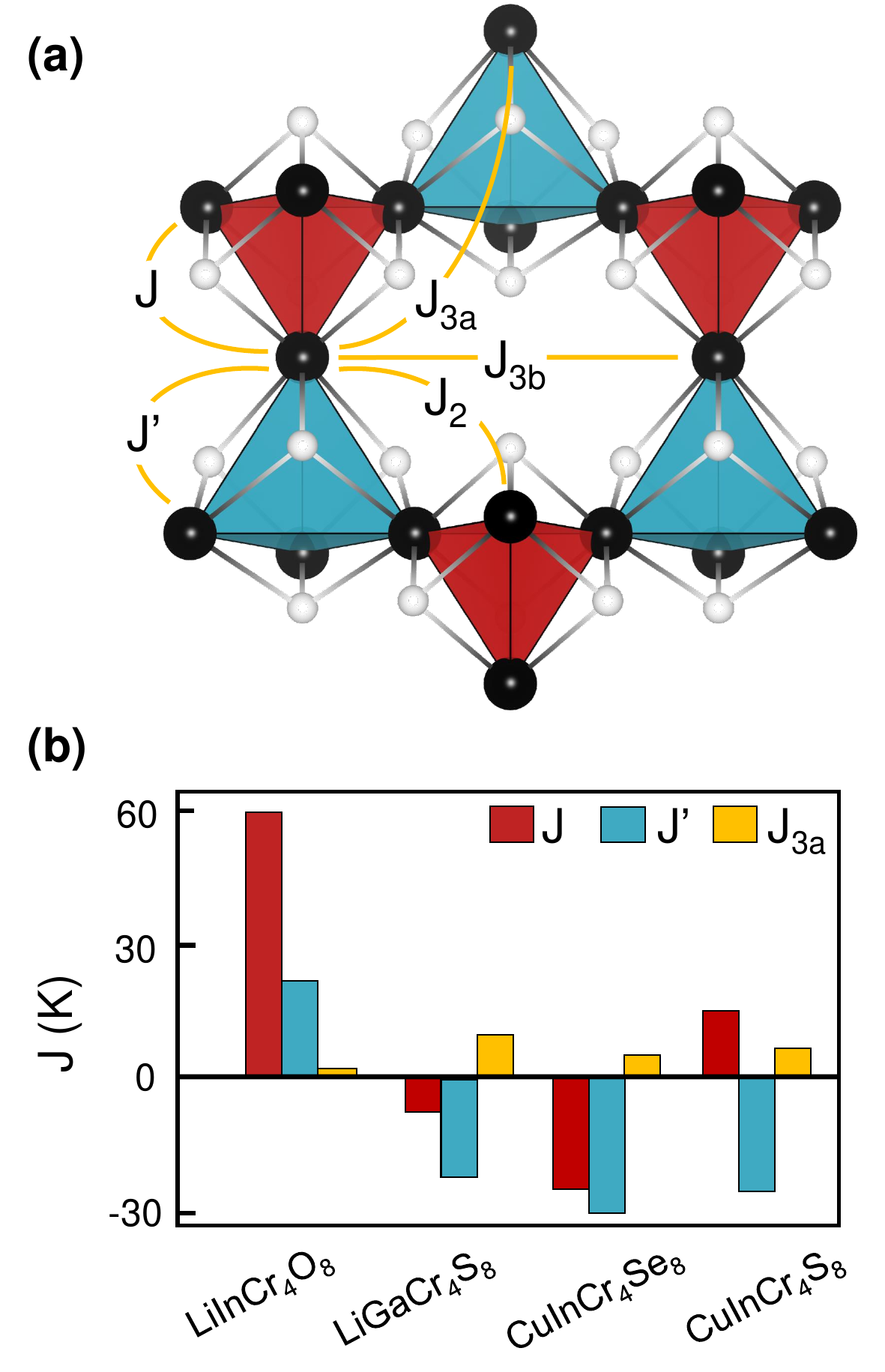}
    \caption{(a) Near neighbor exchange couplings in the breathing pyrochlore chromium spinels, where the chromium atoms are given by the black spheres and the chalcogen is given by the white sphere. The alternating small and large tetrahedra are indicated in red and teal with intratetrahedral exchange couplings $J$ and $J'$, respectively. The second and third nearest neighbor exchange couplings are $J_2$, $J_{3a}$, and $J_{3b}$. (b) This bar graph shows the magnitude and sign of the leading exchange coupling constants, $J$, $J'$, and $J_{3a}$ determined from DFT calculations in Ref.~\cite{19_Ghosh_breathing_Cr_pyrochlore_spinels} for four different breathing pyrochlore compounds. Here positive values of $J$ represent antiferromagnetic couplings and negative values of $J$ are ferromagnetic.}
    \label{fig:breathing}
\end{figure}

In the past few years, there has been a surge of interest in materials with so-called breathing pyrochlore lattices, in which the corner sharing tetrahedral network alternates between large and small tetrahedra. 
The properties of these materials are often understood in terms of their breathing ratios, $B_r = d'/d$ where $d$ and $d'$ are the bond lengths within the small and large tetrahedra, respectively. 
The breathing ratio is a good proxy for the relative strength of their inter-~and intra-tetrahedron exchange couplings, $J$ and $J'$ (Fig.~\ref{fig:breathing}(a)). 
The non-interacting limit occurs when $J \gg J'$, resulting in a scenario of decoupled small tetrahedra, as realized in Ba$_3$Yb$_2$Zn$_5$O$_{11}$ with $B_r = 1.9$, where the ground state appears to be a nonmagnetic singlet~\cite{kimura2014experimental,haku2016low,rau2016anisotropic}. 
Conversely, when the breathing ratio is closer to unity, as is the case for the chromium spinel oxides LiInCr$_4$O$_8$ and LiGaCr$_4$O$_8$ with $B_r \approx 1.05$, intra-tetrahedral interactions are sufficiently strong to yield collective magnetic behavior~\cite{okamoto2013breathing,tanaka2014novel}. 
The expected magnetic ground state in the interacting limit is dictated by the signs of $J$ and $J'$, under the assumption that further neighbour exchange is negligible~\cite{benton2015ground}. 
In LiInCr$_4$O$_8$ and LiGaCr$_4$O$_8$, the intratetrahedral ($J$ and $J'$) exchange couplings are both antiferromagnetic (Fig.~\ref{fig:breathing}(b)). 
These highly frustrated materials exhibit strong spin-lattice coupling and are only able to achieve long range magnetic order after first undergoing a magnetostructural tetragonal lattice distortion~\cite{okamoto2013breathing,nilsen2015complex,lee2016multistage}.

Non-oxide breathing pyrochlores can be found amongst a site-ordered structural derivative of the chromium chalcogenide spinels. 
These materials have the chemical formula $AA'$Cr$_4X_8$ ($A$~= Li, Cu, Ag; $A'=$~Al, Ga, In; $X=$~S, Se) where $A$ and $A'$ are, respectively, monovalent and trivalent nonmagnetic metals. 
For certain combinations of $A$ and $A'$, there is no site ordering and the two cations are randomly distributed over the diamond sublattice of the spinel structure with space group $Fd\bar{3}m$  (\emph{e.g.} Ag$^{+}$/Ga$^{3+}$ with $X=$~S~\cite{pinch1970some,haeuseler1977gitterschwingungsspektren}). 
However, for most combinations of $A$ and $A'$, there is a zinc-blende ordering of the two nonmagnetic metals, each of which forms its own face-centered cubic sublattice, lowering the space group symmetry to $F\bar{4}3m$ (\emph{e.g.} Cu$^{+}$/In$^{3+}$ with $X=$~S, Se~\cite{lotgering1969magnetic,pinch1970some,haeuseler1977gitterschwingungsspektren}), as shown in Figure~\ref{fig:structures}(c), and with crystallographic details provided in Table~\ref{table:crystal_structures}. 
Magnetic susceptibility measurements on these materials find a paramagnetic moment that agrees well with the expected spin-only value ($S=\sfrac{3}{2})$ for Cr$^{3+}$ in a pseudo-octahedral environment~\cite{unger1975magnetic,18_Okamoto_Cr_breathing_pyrochlore_spinel,pokharel2018negative}. 
While the first reports of these compounds date back more than 50 years, interest in their magnetic ground states has recently been reignited.

The replacement of $X=$~O with $X=$~S or Se in the breathing pyrochlore spinels has three major effects (i) ferromagnetic superexchange is enhanced due to the nearly 90$^{\circ}$ Cr--$X$--Cr bond angles, such that either or both $J$ and $J'$ can be ferromagnetic; (ii) the overall magnitude of the $J$ and $J'$ exchange couplings decrease; and (iii) the second and third nearest neighbour exchange couplings ($J_2$, $J_{3a}$, $J_{3b}$) become non-negligible (see Fig.~\ref{fig:breathing}(b))~\cite{okamoto2013breathing,19_Ghosh_breathing_Cr_pyrochlore_spinels,pokharel2020cluster}. 
The net result is that an entirely distinct set of magnetic ground states are expected for these breathing chalcogenide spinels.
Similar to the oxide analogs, spin-lattice coupling remains significant, as evidenced by an extended region of negative thermal expansion in LiGaCr$_4$S$_8$ coinciding with the temperature interval over which magnetic correlations develop~\cite{pokharel2018negative}. 
However, no structural distortions have been detected. 

In terms of the magnetic properties, one particularly interesting compound among the chromium chalcogenide breathing pyrochlores is LiGaCr$_4$S$_8$. 
While the earliest studies of this material suggested that a heat capacity anomaly near 14~K marked an antiferromagnetic ordering transition~\cite{pinch1970some,18_Okamoto_Cr_breathing_pyrochlore_spinel}, more recent studies have found that it is instead associated with a spin freezing transition~\cite{pokharel2018negative}. 
Both DFT calculations and fits to diffuse neutron scattering data reveal that both $J$ and $J'$ for this material are ferromagnetic and that the origin of frustration is strong antiferromagnetic third nearest neighbor coupling, $J_{3a}$ (Fig.~\ref{fig:breathing}(b))~\cite{19_Ghosh_breathing_Cr_pyrochlore_spinels,pokharel2020cluster}. 
The resulting so-called cluster glass state is a form of emergent frustration in which ferromagnetic tetrahedral clusters can be mapped onto an antiferromagnetically-coupled FCC lattice. 

CuInCr$_4$Se$_8$, also with ferromagnetic $J$ and $J'$ has been predicted to possess an incommensurate spiral magnetic order emerging out of a chiral spin liquid state~\cite{19_Ghosh_breathing_Cr_pyrochlore_spinels}. 
Experimentally, significant sample dependence has been observed in this material, with non-stoichiometric samples adopting a spin glass-like state while the magnetic ground state of stoichiometric samples remains unknown~\cite{pinch1970some,wilkinson1976magnetic,duda2008spin}. 
Neutron diffraction has revealed long range antiferromagnetic order in CuInCr$_4$S$_8$~\cite{plumier1971magnetic,plumier1977observation}, CuGaCr$_4$S$_8$~\cite{wilkinson1976magnetic}, and AgInCr$_4$S$_8$~\cite{plumier1971mise}, with the latter two adopting incommensurate spiral order states. 
A number of the $X =$~S breathing pyrochlore spinels have multiple magnetization plateaus with and without hysteresis suggestive of complex phase diagrams due to the delicate balance of competing near and further neighbour interactions~\cite{18_Okamoto_Cr_breathing_pyrochlore_spinel,20_Gen_CuInCr4S8_high_field}.
Growth of any of these compounds in single crystal form would be a significant advancement that would enable a better understanding of their interesting magnetic states.

\section{Future Directions} \label{future}

In this report, we have focused on the families of non-oxide pyrochlore lattice materials that have received the most detailed investigations to date, namely the fluoride pyrochlores, the rare earth chalcogenide spinels, and the breathing pyrochlore chromium chalcogenides. 
We have also briefly touched on a handful of related compounds including Na$_3M$(CO$_3$)$_2$Cl ($M=$~Co, Mn) and FeF$_3$. 
Our focus in this final section shifts to material families that we hope will receive a similar level of investigation in the years to come. 
Some of these are materials that have been synthesized, in some cases decades earlier, but have never received a detailed investigation of their magnetic properties. Others are families of compounds where we believe new materials lurk, remaining to be discovered.

An example of a family that has been long known but has received relatively little attention is the $ABB'$F$_6$ family ($A=$~Group I and $B= 3d$~transition metal), which belong to the same structure type as the so-called $\beta$-pyrochlore oxides, $A$Os$_2$O$_6$ (space group $Fd\bar{3}m$). 
The $\beta$-pyrochlore fluorides have a pyrochlore sublattice that is occupied by a random distribution of the $B^{2+}$ and $B'^{3+}$ atoms. 
The original reports of this family, from 1967, include more than two dozen compounds~\cite{babel1972structure}. 
Of these, only one compound, CsNiCrF$_6$, appears to have had a detailed investigation into its magnetic properties, which revealed Coulomb phase behavior for the magnetic, charge, and lattice degrees of freedom~\cite{harris1995magnetic,zinkin1997short,fennell2019multiple}. 
However, like the disordered $A$-site pyrochlore fluorides, this material enters a frozen glassy state at low temperatures due to chemical disorder. 
There are no known sulfide or selenide analogs of the  $\beta$-pyrochlores but they could conceivably exist with $B=$~Os or some other $5d$ transition metal.

Another interesting path towards discovering new materials is to search for non-oxides analogs of known oxide pyrochlore derivatives.
For example, the ``tripod kagome'' structure, $R_3$Sb$_3A_2$O$_{14}$ (space group $R\bar{3}m$), where $R$ is a magnetic rare earth and $A=$~Mg or Zn, has emerged as a new family of magnetically frustrated materials just within the last few years~\cite{sanders2016re,dun2016magnetic}. 
This structure is obtained by doubling the pyrochlore formula unit and diluting the two pyrochlore sublattices by one quarter.
When viewed along the the [111] crystallographic direction, the pyrochlore structure can be visualized as alternating triangular and kagome layers; this non-magnetic dilution selectively substitutes on the triangular sublattice, leaving the magnetic kagome layers intact. 
Despite the newness of this topic, it already apparent that these tripod kagome materials can exhibit rich magnetic ground states~\cite{paddison2016emergent,scheie2016effective}.
The natural question then is whether this pyrochlore derivative structure could be stabilized with fluoride and chalcogenide anions. 
A fluorinated version of this structure, $M^{2+}_3B^{2+}_3A^{1+}_2$F$_{14}$ ($M =3d$ transition metal, $B =$ Group 2, and $A= $ Group 1), would have no inherent chemical disorder, unlike the fluoride pyrochlores.

Other interesting prospects exist in the domain of mixed anion systems, where the mixture of anions allows different metal valences to be accessed and hence different magnetic states. 
There are, for example, fluorosulfide and oxyfluoride pyrochlores of the type $A_2B_2$F$_6$O and $A_2B_2$F$_6$S wherein the anions are site-ordered such that the one sulfur or one oxygen fully occupies the $48f$ Wyckoff site~\cite{bernard1975hg2m2f6s}. 
An NMR study on Hg$_2$Cu$_2$F$_6$S, a rare example of a Cu$^{2+}$ pyrochlore lattice material, found that the nearest neighbor coupling was on the order of 100 K, yet there is no magnetic order down to 2 K, indicating strong frustration~\cite{07_Kawabata_HgCuFS}. 
Low temperature topochemical anion substitution reactions may prove a promising route to synthesizing new mixed anion pyrochlore lattice materials, as has been shown in oxynitride molybdate pyrochlores~\cite{14_Clark_Molybdate_pyrochlores}.

Finally, it should be emphasized that the materials we have focused our attention on here are all insulators.
While this is a foregone conclusion in the case of the fluoride pyrochlores due to the strongly electronegative fluoride anion, it need not be so in the case of the sulfide and selenide spinels based on transition metals with partially filled $d$-shells. 
One such example is CuIr$_2$S$_4$, which at room temperature is a paramagnetic metal with a pyrochlore lattice that is occupied by a mixture of non-magnetic Ir$^{3+}$ and magnetic Ir$^{4+}$~\cite{furubayashi1994structural,radaelli2002formation}.
At $T_{\rm{MI}} = 230$~K, CuIr$_2$S$_4$ undergoes a metal-to-insulator transition accompanied by a structural distortion that disrupts the perfect pyrochlore sublattice and results in a spin dimerized state, where orbital degrees of freedom and magnetic frustration may play a significant role~\cite{kojima2014magnetic}. 
The chalcogenide spinels are therefore a promising platform to study the interplay of magnetic frustration and itinerant electronic degrees of freedom, which is an emerging topic that is likely to deliver entirely new exotic states.

Non-oxide pyrochlore lattice materials are fertile ground for new and unusual magnetically frustrated states of matter. 
While materials breakthroughs over the past decade have lead to significant progress, and many interesting findings, there is still much more to be done.

\begin{acknowledgments}

This work was supported by the the Natural Sciences and Engineering Research Council of Canada and the CIFAR Azrieli Global Scholars program. 
This research was undertaken thanks in part to funding from the Canada First Research Excellence Fund, Quantum Materials and Future Technologies Program. 
Crystal structure figures were made using VESTA~\cite{cite_vesta}. 

\end{acknowledgments}


%

\end{document}